\documentclass[a4paper, amsfonts, amssymb, amsmath, reprint, showkeys, footinbib, ,twoside,superscriptaddress,floatfix,longbibliography]{revtex4-1}

\usepackage{graphicx}%
\usepackage{dcolumn}%
\usepackage{bm}%
\usepackage{float}
\usepackage{xcolor}
\usepackage{mhchem}
\usepackage{gensymb}

\usepackage{comment}
\usepackage[pdftex, bookmarks, pdffitwindow=false, pdfstartview=FitH, pdfdisplaydoctitle, colorlinks, plainpages=false, pdftitle={},pdfauthor={}, pdfpagelabels, hypertexnames, citecolor={blue!50!black},linkcolor={blue!50!black}, urlcolor={blue!50!black}, pdflang={en}, hyperfootnotes=false, breaklinks]{hyperref}

\usepackage{siunitx}

\newcommand{\silabel}{Supplementary Information}

\newcommand{\avg}[1]{\ensuremath{\left<#1\right>}}

\newcommand{\subfigimg}[3][,]{%
  \setbox1=\hbox{\includegraphics[#1]{#3}}
  \leavevmode\rlap{\usebox1}
  \rlap{\hspace*{0pt}\raisebox{\dimexpr\ht1-2\baselineskip}{#2}}
  \phantom{\usebox1}
}
\newcommand{\figLabel}[1]{\textbf{\textsf{\MakeLowercase{#1}}}}  

\makeatletter
\renewcommand{\@seccntformat}[1]{}
\makeatother

\makeatletter
\def\@bibdataout@aps{
 \immediate\write\@bibdataout{
 @CONTROL{
   apsrev41Control, author="48",editor="1",pages="0",title="0",year="1"
 }}
 \if@filesw
  \immediate\write\@auxout{\string\citation{apsrev41Control}}
 \fi
}
\makeatother

\begin{document}

\preprint{APS/123-QED}

\title{Ion-modulated structure, proton transfer, and capacitance \\in the Pt(111)/water electric double layer}

\author{Xiaoyu Wang}
\affiliation{Department of Chemistry, UC Berkeley, California 94720, United States}

\author{Junmin Chen}
\affiliation{Department of Chemistry, UC Berkeley, California 94720, United States}

\author{Zezhu Zeng}
\affiliation{The Institute of Science and Technology Austria, Am Campus 1, 3400 Klosterneuburg, Austria}

\author{Frederick Stein}
\affiliation{Center for Advanced Systems Understanding (CASUS), Helmholtz Zentrum Dresden-Rossendorf, Germany}

\author{Junho Lim}
\affiliation{Department of Chemistry, Pohang University of Science and Technology (POSTECH), South Korea}
\affiliation{Department of Chemistry, UC Berkeley, California 94720, United States}

\author{Bingqing Cheng}
\email{bingqingcheng@berkeley.edu}
\affiliation{Department of Chemistry, UC Berkeley, California 94720, United States}
\affiliation{The Institute of Science and Technology Austria, Am Campus 1, 3400 Klosterneuburg, Austria}
\affiliation{Materials Sciences Division, Lawrence Berkeley National Laboratory, Berkeley, CA, USA}
\affiliation{Chemical Sciences Division, Lawrence Berkeley National Laboratory, Berkeley, CA, USA}
\affiliation{Bakar Institute of Digital Materials for the Planet, UC Berkeley, California 94720, United States}

\date{\today}%

\begin{abstract}
The electric double layer (EDL) governs electrocatalysis, energy conversion, and storage, yet its atomic structure, capacitance, and reactivity remain elusive.
Here we introduce a machine learning interatomic potential framework that incorporates long-range electrostatics, enabling nanosecond simulations of metal–electrolyte interfaces under applied electric bias with near–quantum-mechanical accuracy.
At the benchmark Pt(111)/water and Pt(111)/aqueous KF electrolyte interfaces, we resolve the molecular structure of the EDL, reveal proton-transfer mechanisms underlying anodic water dissociation and the diffusion of ionic water species, and compute differential capacitance.
We find that the nominally inert K$^+$ and F$^-$ ions, while leaving interfacial water structure largely unchanged, screen bulk fields, slow proton transfer, and generate a prominent capacitance peak near the potential of zero charge.
These results show that ion-specific interactions, which are ignored in mean-field models, are central to capacitance and reactivity, providing a molecular basis for interpreting experiments and designing electrolytes.
\end{abstract}

\maketitle

\section{Introduction}

The interface between an electrode and an electrolyte solution underpins a broad range of electrochemical technologies, including energy storage, electrocatalysis, and chemical manufacturing~\cite{wu2022understanding,jeanmairet2022microscopic,Schott2024}. 
At the heart of this interface lies the electric double layer (EDL), a structured zone of solvent and ions that forms in response to surface charge and applied potentials. 
In the classical Gouy–Chapman–Stern (GCS) picture~\cite{stern1924theorie}, the EDL is described as a combination of a compact (Stern) layer and a diffuse layer of point-charge ions inside a dielectric continuum.
However, mean-field models neglect specific ion–water and ion–surface interactions that govern structure, reactivity, and capacitance at metal–water interfaces~\cite{jeanmairet2022microscopic}. 

While surface-sensitive spectroscopies and indirect electrochemical measurements (e.g., potential of zero charge (PZC), differential capacitance) offer partial insight into the EDL~\cite{Schott2024,zhang2023measurement}, atomic-resolution characterization remains limited. 
Interpreting these measurements requires detailed simulations~\cite{zhang2023measurement,nauman2025electric}, but theoretical modelling of the EDL is notoriously difficult.
The EDL is where quantum mechanics, statistical mechanics, and electrostatics meet.
\emph{ab initio} molecular dynamics (AIMD) based on density functional theory (DFT) with explicit solvation provides accurate atomic interactions but is challenged by the limited size and timescale, and problems in controlling electrode potential or charges under periodic boundary conditions (PBCs)~\cite{gross2019modelling,li2024deciphering,khatib2021nanoscale,le2020molecular}.
Classical empirical force fields allow efficient sampling but struggle with polarizability, charge transfer, and bond-breaking reactions~\cite{scalfi2021molecular}.

Machine learning interatomic potentials (MLIPs)~\cite{Deringer2019,keith2021combining} combine near-DFT accuracy with tractable simulation cost.
However, most MLIPs are short–ranged and omit explicit long–range electrostatics, which are problematic for modelling electrochemical systems and preclude molecular dynamics (MD) simulations under external electric fields.
Recent developments have extended MLIPs to incorporate electrical response. 
For example, deep potential long-range (DPLR) models~\cite{zhang2022deep} employ machine-learned Wannier centers as fixed Gaussian charges to simulate ionic aqueous interfaces~\cite{zhang2024electrical,Zhang2025, Zhu2025Machine}. 
For predicting charges on metal electrodes, integration with the Siepmann–Sprik method~\cite{siepmann1995influence,Zhu2025Machine,dufils2023pinnwall} and machine-learned density models for predicting electronic response~\cite{grisafi2023predicting,feng2025machine} have been proposed.

The Latent Ewald Summation (LES) framework~\cite{Cheng2025Latent,kim2024learning,zhong2025machine,kim2025universal} infers atomic charges and the resulting long-range electrostatics directly from energy and force data. 
The charges can reproduce polarization and Born effective charges (BECs), and be used to drive MD simulations under electric fields~\cite{zhong2025machine}. 
However, as LES determines atomic charges from local atomic environments, it does not capture the long-range charge transfer over the metal surface.  
We therefore adopt a hybrid scheme: LES determines charges in the electrolyte, meanwhile the Siepmann–Sprik model~\cite{siepmann1995influence} solves for the electrode charges that minimize electrostatic energy under an external bias and the influence of electrolyte fields. 
The atomic charges render long-range electrostatics,
and the remaining atomic interactions are captured by a short-ranged MLIP.
This combined framework enables finite-electric-field MD simulations under PBCs, which impose a potential difference~$\Delta\Phi$ between anode and cathode.

As validation, we reproduced classical electrostatic potentials and capacitance profiles for Platinum(111)/water interfaces using a hybrid MLIP trained on SPC/Fw reference data (see Methods). 
We then constructed a MLIP based on the PBE-D3 DFT functional to study the canonical Pt(111)/water interface and its perturbation by adding KF electrolyte. 
Pt(111) is a textbook electrode surface in electrocatalysis~\cite{climent2011thirty,DoblhoffDier2023}, while K$^+$ and F$^-$ are simple, monovalent redox-inactive ions that are commonly considered as inert.
We performed finite-field MD simulations of a periodic Pt(111)/water cell with fixed Pt lattice to resolve the EDL structure, analyzed water dissociation and proton transfer kinetics, and computed the differential capacitance under bias potential. 

\section{Results}

\subsection{Structure of the EDLs}

\begin{figure*}
    \centering
   \includegraphics[width=0.9\textwidth]{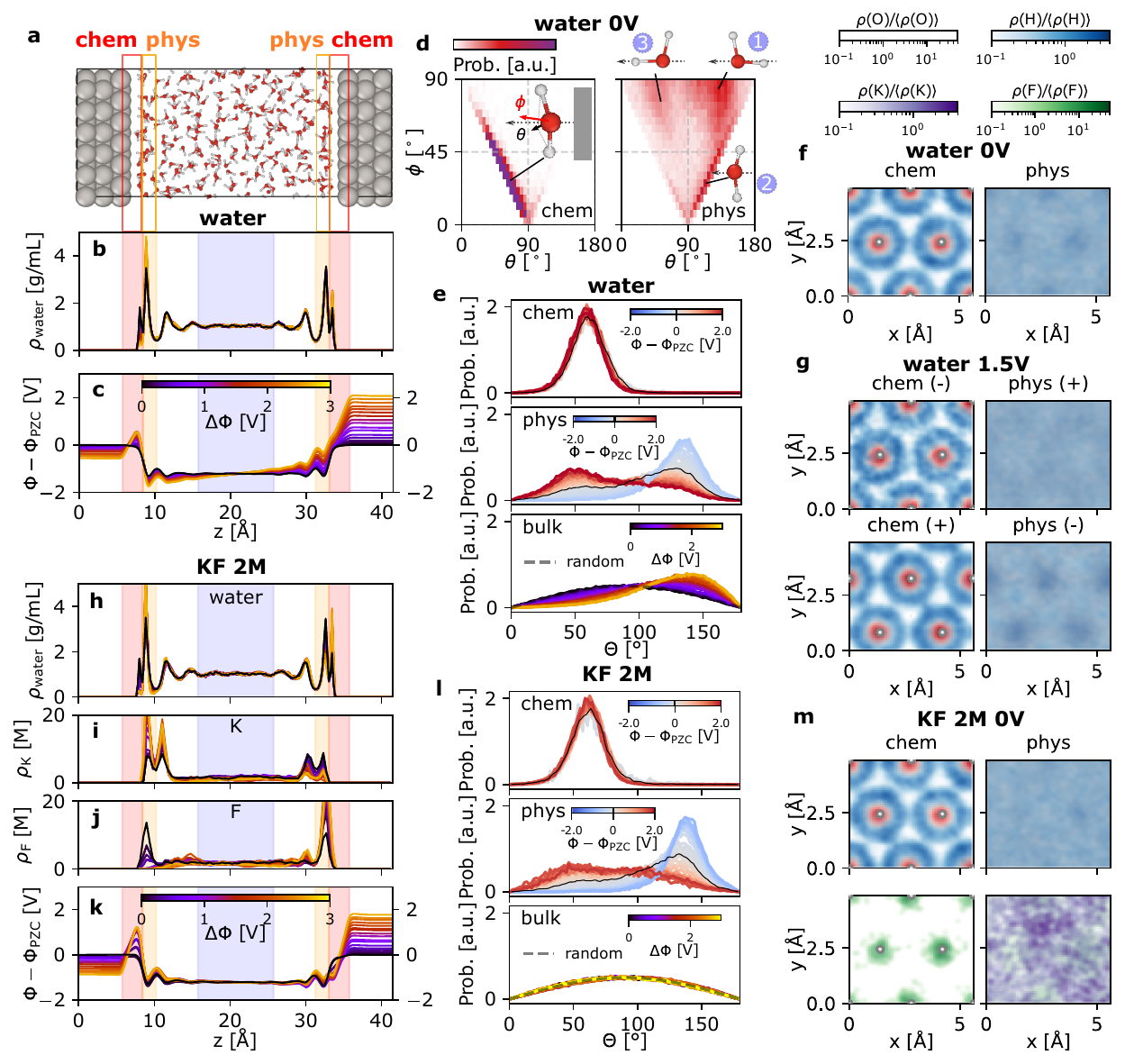}
    \caption{Structure of Pt(111)/water and Pt(111)/aqueous KF 2~M electrolyte electric double layers.\\
    \textbf{a}: Snapshot of Pt(111)/water under zero external electric field. The chemisorbed layers (chem, $<$2.7\AA{} from the surface) and physisorbed layers (phys, 2.7-4.5\AA{} from the surface) are indicated using red and orange boxes, respectively.
    \textbf{b}: The water density profiles $\rho_\mathrm{water}$ along the $z$-axis, perpendicular to the surface. 
    \textbf{c}: Electric potential $\Phi$ along the $z$-axis for Pt(111)/water with the two electrodes having a range of potential difference $\Delta \Phi$. 
    $\Phi=0$ corresponds to the potential of zero charge (PZC) condition.
    \textbf{d}: The orientational distributions of chemisorbed and physisorbed water molecules, from Pt(111)/water under zero external field.
    The inset illustrates that $\theta$ is the angle between the water bisector and the Pt surface normal, and $\phi$ is the angle between the normal vector of the water plane and the surface normal. 
    The most probable water configurations are drawn.
    \textbf{e}: The water dipole angle $\theta$ distribution for chemisorbed, physisorbed, and bulk water molecules in Pt(111)/water under varying $\Delta \Phi$.
    \textbf{f, g, m}: Atomic density maps parallel to the Pt surface at the chemisorbed and physisorbed layers, wrapped around the ($2\times2$) Pt(111) cell. Each panel is normalized by $\avg{\rho}$, the average value of $\rho$ for the corresponding layer. 
    \textbf{f} is for Pt(111)/water under no external field, and \textbf{g} shows the distribution on both the cathode ($-$) and the anode ($+$) under $\Delta\Phi=1.5$~V.
    The shiny gray spheres indicate surface Pt atoms.
    \textbf{h, i, j, k}: The water density profiles $\rho_\mathrm{water}$, K$^+$ ion density $\rho_\mathrm{K}$, F$^-$ ion density $\rho_\mathrm{F}$, electric potential $\Phi$ for the Pt(111)/KF 2~M system under different potential difference $\Delta \Phi$.
    \textbf{l}: The water dipole angle $\theta$ distributions for chemisorbed, physisorbed and bulk water molecules in Pt(111)/KF 2~M.
    \textbf{m}: The atomic density maps for oxygen (red), hydrogen (blue), K (purple), and F (green) for Pt(111)/KF 2~M under no external field.
}
    \label{fig:structure}
\end{figure*}

We begin with the structure of the Pt(111)/water interface under zero bias potential (snapshot in Fig.~\ref{fig:structure}a). 
Water adsorption is characterized by the 
density profiles $\rho (z)$ along the $z$-axis perpendicular to the surface (black curve in Fig.~\ref{fig:structure}b). 
The first peak within 2.7~\AA{} of the Pt surface corresponds to the chemisorbed water (shaded in light red), and the prominent second peak within 2.7~\AA{} to 4.5~\AA{} to the surface corresponds to the physisorbed water (shaded in light orange).
The subsequent peaks are progressively lower in density,
with decaying order up to about 10~\AA{} away from the surface going into the bulk (shaded in light blue).

Both electrodes are charge neutral, 
as they are symmetric and the total charge on the metallic atoms is zero in the Siepmann-Sprik model used.
Accordingly, the potential on either electrode is the PZC.
The electrostatic potential $\Phi(z)$ (black curve in Fig.~\ref{fig:structure}c) referenced to this PZC, 
is determined by solving one-dimensional Poisson's equation based on Gaussian smeared charge density.
$\Phi(z)$ drops by about $1.2$~V from the electrode to the bulk liquid, with the steepest change at the physisorbed layer, consistent with a compact interfacial region bearing most of the potential change.

The orientations of interfacial water are characterized by the dipole angle ($\theta$) between the bisector and the surface normal, and the molecular plane angle ($\phi$) between the water plane normal and the surface normal (Fig.~\ref{fig:structure}d,e). 
Chemisorbed waters adopt a narrow distribution: their molecular planes lie nearly parallel to the surface ($\phi \approx20^\circ$), while their dipoles tilt away ($\theta \approx 60^\circ$) with both H atoms pointing outward. 
Physisorbed waters are more flexible but still anisotropic.
The most probable configuration is water with a near-vertical molecular plane and one H pointing to the surface ($\phi\approx 90\degree$, $\theta \approx 130\degree$), followed by near-flat water with two H atoms pointing towards the surface ($\phi\approx 30\degree$, $\theta \approx 120\degree$). 
There is also a smaller population of molecules with nearly vertical water planes and dipole pointing away ($\phi\approx 80\degree$, $\theta \approx 50\degree$). As we will later discuss, such water orientations play a part in proton transfer events.
By contrast, bulk waters are randomly oriented, with $\theta$ distributions matching the $\sin(\theta)$ baseline.
Water at different layers also have distinct diffusion coefficients, as shown in Fig.S5 of \silabel.

Fig.~\ref{fig:structure}f shows the
in-plane density maps of oxygen and hydrogen atoms, normalized by the layer averages ($\avg{\rho}$) of the specific atom types.
In the chemisorbed layer, O atoms sit directly on top of surface Pt atoms (gray spheres), while hydrogens form ring–like features around those sites due to the predominantly in–plane molecular orientation.
Not all Pt atoms adsorb oxygen; the surface coverage is about 15\% under no bias potential.
For the physisorbed layer, such a pattern is completely lost, and no obvious ordering can be found.

We then impose a potential difference $\Delta\Phi$ between the cathode ($-$, left) and anode ($+$, right) electrodes by applying a uniform electric field of magnitude $\Delta\Phi/l_z$ along the $z$-axis. 
The potential profiles $\Phi(z)$ are shown as the colored curves in Fig.~\ref{fig:structure}c, and $\Phi(z)$ collected at different $\Delta\Phi$ are aligned by assuming that the average potential of the bulk region is the same. 
The gradient of $\Phi(z)$ is rather localized near the interface, with the most rapid change in the physisorbed layer in the anode.
The change of the potentials away from the PZC in the cathode (left) and the anode (right) is not symmetric, with more changes in the anode. 
In the bulk region, $\Phi(z)$ are nearly linear with a small slope, indicating a constant electric field that is smaller than the applied external field.

The colored curves in Fig.~\ref{fig:structure}b show the water density across the interface under different $\Delta \Phi$ values.
On the cathode, chemisorbed water decreases in population while the physisorbed peak grows with increasing bias. The anode exhibits the opposite trend. 
Densities deeper in the liquid remain unchanged, confirming that the electric field primarily affects the interfacial region.
The influence of $\Delta \Phi$ on the diffusion coefficients is shown in Fig.S5 of \silabel.

The bias-dependent orientational response is shown in Fig.~\ref{fig:structure}e. 
Chemisorbed waters remain rigid, with dipole orientations nearly unaffected by the bias. 
Physisorbed waters are more flexible, tilting to higher $\theta$ at the cathode and lower $\theta$ at the anode. 
Bulk water also shows a dipole alignment consistent with the finite residual field in the slab interior (Fig.~\ref{fig:structure}c).

In-plane atomic density maps (Fig.~\ref{fig:structure}g) under $\Delta \Phi=1.5$~V are largely similar to the zero bias case:
ordered chemisorbed layers and disordered physisorbed layers at both the anode and the cathode.
For the chemisorbed layer, the cathode density map has more statistical noise due to fewer chemisorbed water (Fig.~\ref{fig:structure}b).
The H density on the anode shows bridging sites between the blue rings, which we will explain later.

Next, we probe ion effects on the EDL by simulating the same Pt(111) electrode in an aqueous KF electrolyte at 2~M. 
The water density profiles $\rho_\mathrm{water}(z)$ in Fig.~\ref{fig:structure}h closely resemble the pure-water case, suggesting that finite ionic strength does not qualitatively alter interfacial layering of water. 
Under applied potential difference $\Delta\Phi$, the redistribution between the chemisorbed and physisorbed populations follows the same dependence observed without electrolyte, with a slightly larger shift.

The K and F atom number density profiles (Figs.~\ref{fig:structure}i,j) exhibit pronounced accumulation near the surface even at zero potential bias: 
K$^+$ displays two maxima at about 3.3~\AA{} and 5.2~\AA, residing in the physisorbed layer and the subsequent solvent layer. 
F$^-$ adsorption is closer to the surface; a dominant peak centers at around 3.1~\AA{} from the surface, largely within the physisorbed layer but with pronounced weight extending into the chemisorbed layer.
As $\Delta \Phi$ increases, K$^+$ accumulates at the cathode and F$^-$ at the anode,
dramatically adding to the height of the adsorption peaks.

The $\Phi(z)$ profiles for Pt(111)/KF 2~M (Fig.~\ref{fig:structure}k) contrast sharply with water (Fig.~\ref{fig:structure}c). 
In the electrolyte, $\Phi(z)$ exhibits an extended, nearly flat plateau in the slab interior, indicating a negligible residual field in the bulk. 
This is because ion redistribution screens the external field, concentrating the potential drop within the surface layers. 
Such strong screening is consistent with the classic theory that assigns the Debye length of $\approx 2$~\AA{} for the 2M ionic strength.
Consequently, bulk water remains essentially unpolarized: the $\theta$–distributions fall onto the random baseline at all $\Delta\Phi$ (Fig.~\ref{fig:structure}l). 
This contrasts with the polarized bulk water in the Pt(111)/water case (Fig.~\ref{fig:structure}e).
Meanwhile, the orientational response of the interfacial water in the Pt(111)/KF 2~M (Fig.~\ref{fig:structure}l) mirrors the trends observed for pure water: chemisorbed waters are rigid, and physisorbed waters reorient under bias. 
The in-plane density maps of oxygen and hydrogen atoms (Fig.~\ref{fig:structure}m) are also similar to the water case (Fig.~\ref{fig:structure}f).
In addition, F$^-$ ions show specific bonding with Pt atoms, similar to what is observed for O atoms.
K$^+$ ions exhibit no notable in-plane ordering.

In summary, even though the added KF ions have largely accumulated at the interface area, they produce only small, quantitative changes to interfacial water structure in the Pt(111)/aqueous EDLs.
The dominant effect is electrostatic screening: the potential drop is confined to the compact interfacial layers and the bulk electric field vanishes, leaving the bulk water unpolarized.

\subsection{Water dissociation and proton transfer reactions}

\begin{figure*}
    \centering
   \includegraphics[width=0.8\textwidth]{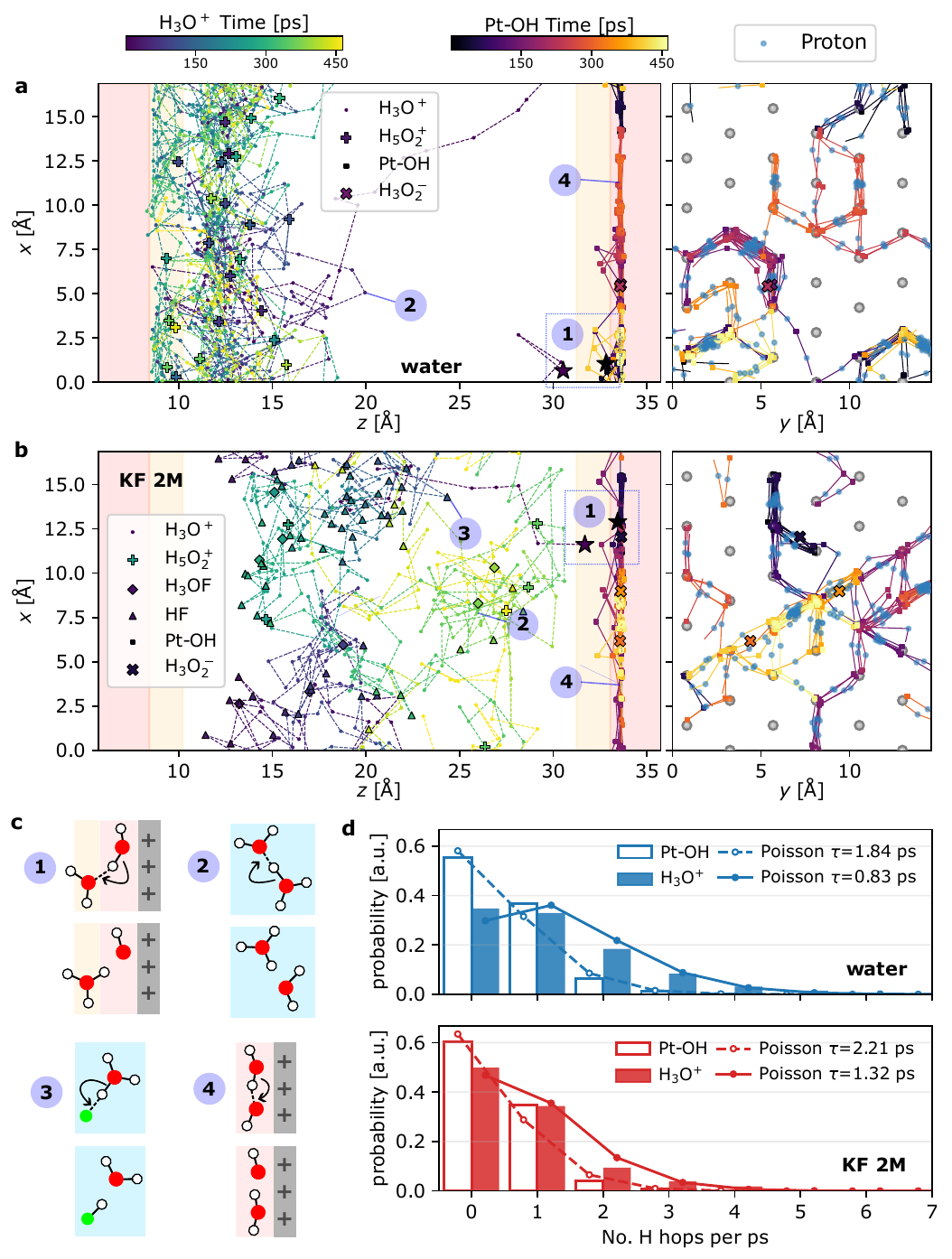}
    \caption{ Proton-transfer reactions at Pt(111)/water electric double layers with and without KF electrolyte.\\
\textbf{a}: Time evolution of proton-transfer events in the Pt(111)/water system under $\Delta \Phi = 1.88$~V.  
The left panel ($x$–$z$ projection) shows the positions of participating species (defined by the mean positions of O and/or F atoms) since the onset of water dissociation. Red and orange shaded regions indicate the chemisorbed and physisorbed layers, respectively.  
The right panel ($x$–$y$ projection) shows the trajectories of Pt-OH together with the hopping protons (blue circles). 
Surface Pt atoms are shown as shiny gray spheres.
\textbf{b}: Time evolution of proton-transfer events in the Pt(111)/KF 2~M system under $\Delta \Phi = 1.65$~V.  
\textbf{c}: Schematic of four proton-hopping mechanisms: (1) proton donation from a chemisorbed water molecule to a physisorbed water at the anode ($+$), (2) Grotthuss-type proton hopping between water molecules, (3) proton transfer between hydronium and F$^-$ forming a transient HF, and (4) hydroxide-mediated hopping along the chemisorbed layer via inverse Grotthuss steps at the anode ($+$).
\textbf{d}: Distribution of proton-hopping events per picosecond from simulations (bars), compared with Poisson distributions (lines) derived from the mean hopping time $\tau$ for Pt(111)/water and Pt(111)/KF 2~M.
}
    \label{fig:reaction}
\end{figure*}

We observe anodic water dissociation events forming surface hydroxyls (Pt-OH) and solvated hydroniums (H$_3$O$^{+}$) during our nanosecond-scale MD simulations, which start with only molecular water. 
These dissociation events are rare and activated;
our trajectories do not provide ergodic sampling of dissociation rates or equilibrium concentrations. 
A full list of events is summarized in the \silabel. In Pt(111)/water, higher applied potential $\Delta \Phi$ seems to promote dissociation: under $\Delta \Phi \geq 1.5$~V persistent dissociation appears within nanoseconds, though with variable onset times,
whereas only a couple of short-lived dissociation events under lower $\Delta \Phi$ occurred before recombination happened.
Here, we focus on the proton-transfer mechanisms. 
Representative trajectories are shown in Fig.~\ref{fig:reaction}a for Pt(111)/water under $\Delta \Phi = 1.88$~V and Fig.~\ref{fig:reaction}b for Pt(111)/KF 2~M under $\Delta \Phi = 1.65$~V.

The formation of an H$_3$O$^{+}$ and surface Pt–OH pair at the anode (black stars in Fig.~\ref{fig:reaction}a,b) proceeds via proton transfer from a chemisorbed water (Pt–H$_2$O) to a nearby physisorbed water (mechanism 1 in Fig.~\ref{fig:reaction}c). 
The donor water lies nearly in-plane with both hydrogens pointing away from the surface, 
while the acceptor is nearly vertical with its oxygen directed toward the surface (configuration type 3 in Fig.~\ref{fig:structure}d). 
This geometry shortens the donor–acceptor H–O distance, facilitating transfer.

The resulting H$_3$O$^{+}$ then diffuses away via the Grotthuss mechanism, relaying its excess proton through hydrogen bonds with neighboring water molecules (mechanism 2 in Fig.~\ref{fig:reaction}c). 
This Grotthuss mechanism agrees with previous AIMD and reactive force fields studies on bulk water~\cite{knight2012curious,tse2015analysis}.
We observe transient complexes such as Zundel cations (H$_5$O$_2^+$) and H$_3$OF, which act as short-lived intermediates in proton transfer.
Overall, Pt(111)/water and Pt/KF 2~M share common proton-transfer pathways, with the ions introducing an additional channel involving the formation of HF from proton transfer to F$^-$ (mechanism 3).
Spatial distributions of the hydronium differ: in pure water, H$_3$O$^+$ localizes near the physisorbed layer at the cathode, driven by the residual field (Fig.~\ref{fig:structure}c), whereas in KF 2~M, electrolyte screening yields a broader, more uniform distribution of hydronium.

Meanwhile, Pt–OH remains confined to the chemisorbed layer but diffuses laterally via inverse-Grotthuss steps: 
a proton can hop from a chemisorbed H$_2$O to a neighboring Pt–OH species (mechanism 4 in Fig.~\ref{fig:reaction}c).
As shown in the in-plane trajectory of Pt–OH (Fig.~\ref{fig:reaction}a, right panel),
a hopping proton (blue circles) sits near the midpoint of two O chemisorbed atoms,
with the intermediate state being the hydroxide-water complex (crosses).
This explains the bridging features between the hydrogen density rings in Fig.~\ref{fig:structure}g.
There are frequent protons hopping back and forth, yielding long correlation times in the propagation of Pt–OH positions.

We analyze proton transport dynamics from MD trajectories where water dissociation occurred.
Since hopping rates showed no clear dependence on $\Delta \Phi$ (see \silabel), 
we aggregate data across different biases in Fig.~\ref{fig:reaction}d.
The mean hopping time $\tau$ is defined as the trajectory duration with one dissociated water molecule divided by the number of proton hops.
For water, $\tau$ is 0.83~ps for H$_3$O$^+$ and 1.84~ps for Pt-OH, indicating the faster mobility of the protonated species.
In KF 2~M, $\tau$ increases to 1.32~ps (+60\%) for H$_3$O$^+$ and 2.21~ps (+20\%) for Pt-OH.
The H$_3$O$^+$ hopping times align with experimental estimates ($\sim$1~ps)~\cite{meiboom1961nuclear,woutersen2006ultrafast,Yuan2019}, and reproduce the experimentally observed slowdown in acids~\cite{Yuan2019}.
In addition, the distributions of hopping events per picosecond for both systems (bar plots) are well described by Poisson statistics (solid lines) parameterized by the $\tau$, suggesting that each proton transfer can be considered a stochastic and independent event at the 1~ps timescale.

\subsection{Differential capacitance and the origin}

\begin{figure*}
  \centering
\includegraphics[width=0.8\textwidth]{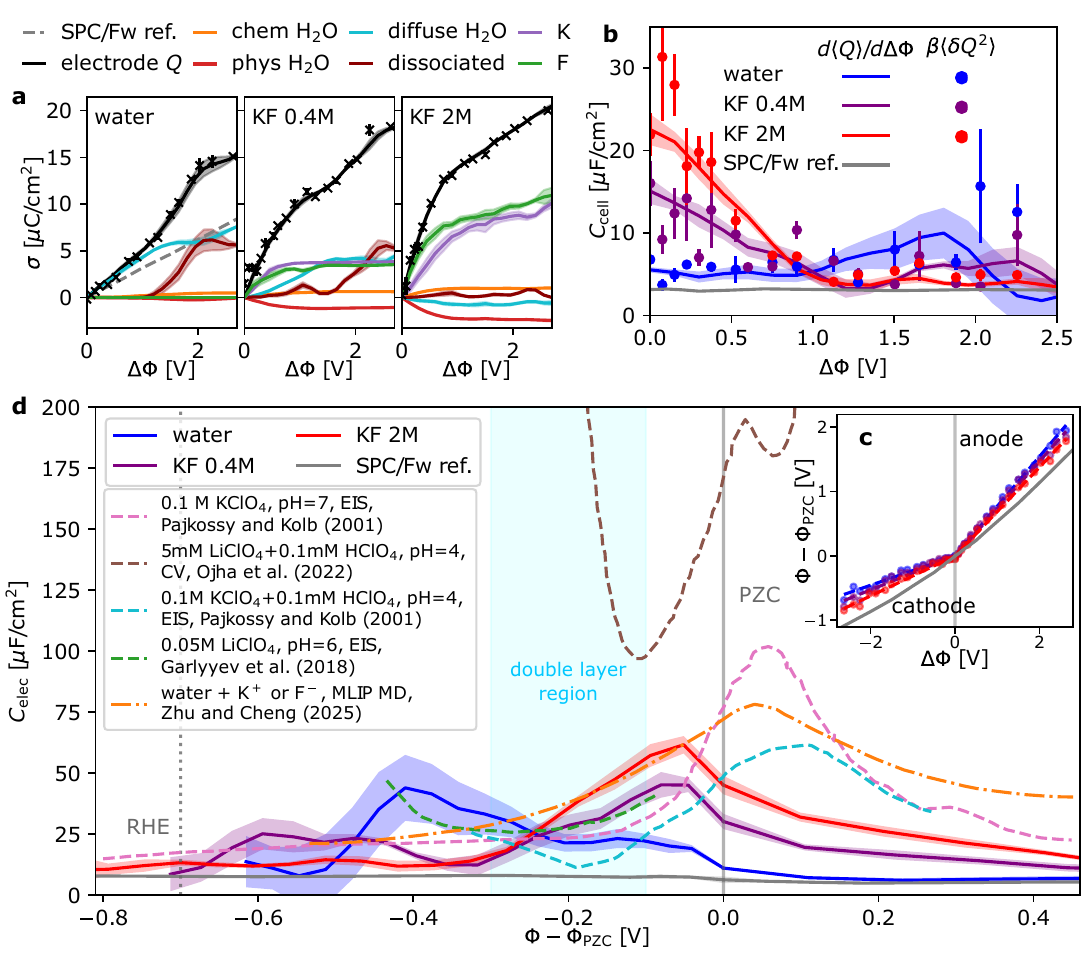}
  \caption{
  Differential capacitance of Pt(111)/water and Pt(111)/aqueous KF electrolyte interfaces.\\
  \textbf{a}: Area–specific charge $\sigma=\avg{Q}/S$ for Pt(111)/water (left), Pt(111)/KF 0.4~M (middle), and Pt(111)/KF 2~M (right) under different electrode potential difference $\Delta \Phi$, obtained from the reweighting estimator in Eqn.~\eqref{eq:reweight}; shaded bands indicate uncertainties. 
  Crosses with error bars are independent averages from simulations run under the corresponding $\Delta\Phi$. 
  The gray curve is from finite–field simulations using classical SPC/Fw water. 
  The colored lines show area-specific polarization divided by the distance between two electrodes, $P/\Delta l$, 
  from different contributions: undissociated chemisorbed water (chem H$_2$O), 
  undissociated physisorbed water (phys H$_2$O), 
  undissociated water from the remaining region (diffuse H$_2$O), 
  dissociated water.
\textbf{b}: Cell differential capacitance $C_{\mathrm{cell}}$ with respect to $\Delta \Phi$, computed as both the numerical derivative $d\langle Q \rangle/d\Delta\Phi$ (lines with bands) and from charge fluctuations $\beta\langle \delta Q^2\rangle$ (symbols with error bars). 
The gray curve is the classical reference using SPC/Fw water.
\textbf{c}: Electrode potentials relative to the potential of zero charge (PZC) as functions of $\Delta \Phi$. 
\textbf{d}: Electrode differential capacitance $C_{\mathrm{elec}}$ as a function of $\Phi - \Phi_\mathrm{PZC}$. 
The MLIP MD results from Zhu and Cheng~\cite{Zhu2025Machine} are for a compact electric double layer at the high
concentration limit.
Experimental data are included for comparison (Pajkossy and Kolb~\cite{pajkossy2001double}, Ojha et al.~\cite{ojha2022double}, Garlyyev et al.~\cite{garlyyev2018influence}), with experimental conditions (technique: cyclic voltammetry (CV) or electrochemical impedance spectroscopy (EIS), pH, electrolyte composition) indicated in the legend. 
For experiments, PZC values are calculated as $0.29 + 0.059\,\mathrm{pH}$~V$_\mathrm{RHE}$. 
Reversible hydrogen electrode (RHE) at pH=7 and PZC are indicated using dotted and solid gray vertical lines, respectively. 
The experimental double layer region (0.4-0.6~V vs RHE~\cite{ojha2020double}) is marked by a light cyan shade.
  }
    \label{fig:C}
\end{figure*}

In finite-field MD with a potential difference $\Delta\Phi$ across the two Pt(111) electrodes, each electrode develops a charge density $\sigma=\langle Q\rangle/S$, where $\langle Q\rangle$ is the mean electrode charge and $S$ the surface area.
Figure~\ref{fig:C}a shows $\sigma$ for Pt(111)/water, Pt(111)/KF 0.4~M, and Pt(111)/KF 2~M.
The reweighting scheme (black curves with uncertainty bands, see Methods) yields smooth, low-variance estimates interpolating the direct point averages (crosses).
Compared to water, KF electrolytes show consistently larger $\sigma$ and a steeper slope near zero bias.

We rationalize the charge response by decomposing the electrolyte polarization $P$. 
In the parallel–plate slab geometry, the conducting metal surface charge density satisfies
$\sigma = P/\Delta l + C_{\mathrm{empty}} \Delta\Phi$,
where $\Delta l$ is the electrode separation and $C_{\mathrm{empty}}$ the vacuum capacitance of the bare electrodes.
We partition $P$ into contributions from undissociated water in the chemisorbed, physisorbed, and diffuse regions, as well as from K$^+$, F$^-$, and (when present) dissociated water.
These $P/\Delta l$ components are shown in Fig.~\ref{fig:C}a in the same units as $\sigma$.

In Pt(111)/water, undissociated water in the chemisorbed and physisorbed layers contributes very little to the net polarization $P$; most of the response arises from the remaining diffuse region. 
The $P(\Delta\Phi)$ values of such undissociated water closely match the classical simulations that employ SPC/Fw water, Pt(111) described by the Siepmann-Sprik model and Lennard-Jones water-metal interactions (see Methods). 
For $\Delta\Phi \gtrsim 1.5$~V, water dissociation emerges, and the resulting species add a growing contribution to $P$, 
although this may be underestimated as the equilibrium concentrations of H$_3$O$^+$ and Pt–OH are not converged.

In Pt(111)/KF 2~M, the electrolyte polarization is dominated by the ions: 
K$^+$ and F$^-$ account for the vast majority of $P$, whereas water contributes a smaller component of opposite sign. 
Diffuse water contributes little; most of the water response arises from the physisorbed layer. 
This can be understood as the bulk water molecules remain randomly distributed as the ions completely screen the electric field in the bulk (as shown in Fig.~\ref{fig:structure}k and l).
The water dissociation $P$ is more subtle compared to the pure water case, due to the smaller separation between H$_3$O$^+$ and Pt-OH at the absence of residual bulk electric field (Fig.~\ref{fig:reaction}b).
In Pt(111)/KF 0.4~M, the response at low-bias ($\Delta\Phi \lesssim 0.5$~V) resembles the 2~M case: polarization is dominated by K$^+$ and F$^-$ ions. 
At higher $\Delta\Phi$, the only 2 ion pairs in the simulation box are all adsorbed on the surfaces and exhausted in the bulk; 
There is a residual electric field in the diffuse region similar to the pure water case, and correspondingly, the polarizations of diffuse water and dissociated water grow.

The differential capacitance of the simulation cell can be computed as the derivative of the electrode surface charge density with respect to $\Delta\Phi$:
\begin{equation}
    C_{\mathrm{cell}} = \dfrac{d \avg{\sigma}}{d \Delta \Phi}
    = \dfrac{1}{S}\dfrac{1}{k_\mathrm{B} T} \avg{\delta Q^2}.
    \label{eq:cdiff}
\end{equation}
Alternatively, based on the fluctuation-dissipation relation, 
it can also be derived from the variance of the total charge distribution~\cite{limmer2013charge,scalfi2020charge} 
with $\delta Q = Q - \langle Q \rangle$.

Figure~\ref{fig:C}b shows that capacitances $C_\mathrm{cell}$ from numerical derivatives of $\sigma(\Delta\Phi)$ (lines with uncertainty bands) agree with those from charge fluctuations, $\beta\langle\delta Q^2\rangle$ (symbols), confirming the statistical consistency of the finite-field MLIP MD simulations.
For Pt(111)/water, $C_{\mathrm{cell}}$ is $5.5(3)~\mu\mathrm{F/cm}^2$ at zero bias and remains nearly flat up to $\Delta\Phi \lesssim 1.5$~V. 
A broad peak appears at $\Delta\Phi \approx 1.7$~V, arising from polarization of dissociated water (dark red curve in Fig.~\ref{fig:C}a), though its height may be underestimated.
Classical Pt(111)/SPC/Fw simulations show a similar trend but with a smaller baseline ($3.1(7)~\mu\mathrm{F/cm}^2$ at zero bias) and no significant rise at higher voltages.
In both cases, $C_\mathrm{cell}$ is far below the continuum estimate of $\sim23~\mu\mathrm{F/cm}^2$ based on the bulk water dielectric constant ($\varepsilon\approx78$) and the 30~\AA{} slab thickness. This suppression, consistent with previous classical MD studies, reflects restricted orientational freedom and saturation of interfacial dipoles that limit bulk-like dielectric screening at the interface~\cite{limmer2013charge,Limaye2024,nauman2025electric}.

Upon the addition of KF electrolyte, the zero-bias $C_{\mathrm{cell}}$ rises sharply: Pt(111)/KF 0.4~M peaks at $15(2)~\mu\mathrm{F/cm}^2$ and Pt(111)/KF 2~M reaches $23(2)~\mu\mathrm{F/cm}^2$. 
At higher $\Delta\Phi$, $C_{\mathrm{cell}}$ decreases toward the pure water baseline.
Pt(111)/KF 0.4~M also shows a broad peak at $\Delta\Phi \approx 2.2$~V due to the dissociated water contribution.
Relating back to the polarization decomposition in Fig.~\ref{fig:C}a, the capacitance enhancement at $\Delta\Phi\approx0$~V originates from the strong polarization and adsorption of K$^+$ and F$^-$ ions on the oppositely charged electrodes, which screen the applied field more effectively than water dipoles alone. 

The single–electrode differential capacitance, typically measured experimentally, is given by $C_{\mathrm{elec}} = d\langle\sigma\rangle/d\Phi$, where $\Phi$ is the potential of the electrode. 
Its relation to the cell capacitance is $1/C_{\mathrm{cell}} = 1/C_{\mathrm{anode}} + 1/C_{\mathrm{cathode}}$.
Figure~\ref{fig:C}c shows how $\Delta\Phi$ maps onto electrode potentials relative to the PZC. 
As seen from $\Phi(z)$ (Fig.~\ref{fig:structure}c), the potential changes more steeply at the anode than the cathode due to asymmetric water structures and/or ion adsorption. 
Consequently, $C_{\mathrm{elec}}$ is larger at negative $\Phi-\Phi_\mathrm{PZC}$ (cathode) than at positive values (anode) (Fig.~\ref{fig:C}d).

The $C_{\mathrm{cell}}$ peak for Pt(111)/water appears near $\Phi=-0.4$ V ($\Delta\Phi \approx 1.7$ V), driven by anodic water dissociation that yields surface Pt–OH and solvated H$_3$O$^+$.
A similar dissociation peak occurs around $\Phi=-0.6$ V for Pt(111)/KF 0.4 M.
Otherwise, $C_{\mathrm{elec}}$ for pure water near the PZC is low and close to the SPC/Fw reference, whereas both Pt(111)/KF 0.4 M and 2 M show a pronounced peak at the PZC, reflecting the $C_{\mathrm{cell}}$ maximum at zero bias.

We discuss the limitations of these computed capacitances:
First, the magnitude of the water dissociation peak is likely underestimated due to limited sampling of rare reaction events.
Second, our simulations neglect surface H adsorption, which is expected below $\Phi \lesssim 0.4$ V versus the reversible hydrogen electrode (RHE)~\cite{le2021modeling}.
In other words, our capacitances account for the polarization of molecular water and ions, but largely omit the  contributions from H and OH surface adsorption.
Thus, our $C_{\mathrm{cell}}$ estimates are most reliable in the double layer region of the Pt(111)/water interface (estimated to be 0.4–0.6 V vs RHE~\cite{ojha2020double}, shaded area in Figure~\ref{fig:C}d), where Faradaic processes and specific water ion adsorption are minimal.

Our predicted $C_{\mathrm{elec}}$ curves for Pt(111)/KF electrolytes qualitatively resemble previous AIMD and MLIP studies~\cite{le2020molecular,Zhu2025Machine}, 
which also report a broad capacitance peak near the PZC. 
However, those works~\cite{le2020molecular,Zhu2025Machine} induced electrode charge differently: only one electrode was modeled, and fixed counterions (e.g., K$^+$ or F$^-$) were placed near the surface to enforce charge neutrality. Such a setup models a compact electric double layer at the high concentration limit, although the de facto electrolyte concentration is ambiguous. Moreover, the specific ways to implement such counter charge methods can affect the capacitance estimates~\cite{li2024deciphering}.

Fig.~\ref{fig:C}d also compares four curves of $C_{\mathrm{elec}}$ in near pH neutral electrolytes measured using different experimental techniques~\cite{pajkossy2001double,ojha2022double,garlyyev2018influence}.
Note that the PZC values were calculated as $0.29 + 0.059\,\mathrm{pH}$~V$_\mathrm{RHE}$~\cite{zhang2023measurement}, although there is uncertainty~\cite{cuesta2004measurement}.
In general, no agreement can be found between the experimental results, and all of them defy the GCS prediction of a capacitance minimum at the PZC.
The discrepancies were attributed to the differences in the
electrolyte composition, cleanliness, experimental techniques, and the model assumptions used to extract $C_{\mathrm{elec}}$ from raw data~\cite{zhang2023measurement}.
Our estimated $C_{\mathrm{elec}}$ values in the double layer region fall within the experimental range extracted from electrochemical impedance spectroscopy (EIS)~\cite{pajkossy2001double,garlyyev2018influence}, where equivalent circuit analysis minimizes contributions from faradaic processes and specific adsorption. 
In contrast, capacitances from cyclic voltammetry (CV)~\cite{ojha2022double}, which include charging, adsorption, and redox currents, are significantly larger.
The strong dependence of our estimated capacitance on ion concentration and anodic water dissociation may help interpret these experiments, particularly for the observed ion concentration dependence~\cite{DoblhoffDier2023,ojha2022double}.

\section{Conclusions}

We develop a long–range electrostatic machine-learning interatomic-potential framework that couples Latent Ewald Summation~\cite{Cheng2025Latent,kim2024learning,zhong2025machine} with the Siepmann–Sprik metal model~\cite{siepmann1995influence}, enabling near-DFT-accuracy atomistic simulations of electrode–electrolyte interfaces under applied potential difference. 
Compared to the other MLIPs methods that can be coupled to electric fields~\cite{zhang2022deep,Zhang2025,feng2025machine}, this framework has higher training accuracy while only utilizing the standard energy and force training labels, without training on Wannier centers or the finicky finite-field DFT calculations. 
We resolve the interfacial structure, anodic water dissociation and proton-transfer pathways, as well as differential capacitance at the Pt(111)/water interface, then assess the influence of redox-inactive ions (K$^+$ and F$^-$) on these properties.

On the structures (Fig.~\ref{fig:structure}), K$^+$ and F$^-$ accumulate at the interface, with F$^-$ approaching closer to (and binding at) Pt sites. 
This is consistent with a Stern-layer picture, but beyond what mean-field GCS can predict about ion specificity. 
Despite this adsorption, KF leaves interfacial water adsorption and orientational distributions largely intact. 
Under external electric potential, ionic redistribution screens the field in the bulk and localizes the potential drop within the compact interfacial region. 

On the reactivity and dynamics, both pure water and KF electrolyte share the same network of proton–transfer pathways spanning chemisorbed, physisorbed, and diffuse layers: anodic water dissociation forming Pt-OH and H$_3$O$^+$ via proton donation from
a chemisorbed water to a physisorbed water (mechanism 1 in Fig.~\ref{fig:reaction}c),
the Grotthuss mechanism for H$_3$O$^+$ migration (mechanism 2),
and the inverse Grotthuss mechanism for Pt-OH confined to the chemisorbed layer (mechanism 4).
The addition of F$^{-}$ enables another proton transfer channel via transient HF formation (mechanism 3).
The added ions reduce proton-hopping frequency for both Pt-OH and H$_3$O$^+$ diffusion.

Ion-driven polarization drastically changes the capacitance behavior of the Pt(111)/aqueous KF electrolyte compared to the pure water case.
Notably, the electrolytes display a strongly enhanced $C_{\mathrm{elec}}$ near the PZC, while the capacitance peak related to water dissociation is less pronounced (Fig.~\ref{fig:C}c).
It should be noted that our calculations likely underestimate the contribution coming from the Pt-OH and H$_3$O$^+$ species, and completely omit surface H.
Such specific H and OH adsorptions may be important outside the double layer region~\cite{ojha2020double}, and will be investigated in the future.

Our results link the specific ion–water, ion–surface, and water-surface interactions with the macroscopic behaviors of the Pt(111)/water EDL.
In particular,  
we demonstrate that inert ions such as K$^+$ and F$^-$ actively reshape proton pathways and capacitance by screening fields and introducing ion-specific interactions. 
These atomistic insights help explain the experimentally observed sensitivity of capacitance to electrolyte concentration and identity~\cite{DoblhoffDier2023,ojha2022double}, and provide guidance for electrolyte design.

The long–range MLIP framework developed here is applicable to study a wide range of interfaces across energy storage, conversion, and catalytic systems, such as
nanocapacitors, hydrogen and oxygen evolution reactions, CO$_2$ reduction, and solid–electrolyte interphase.
By combining quantum-mechanical accuracy with nanosecond-scale sampling under finite electric fields, it opens the door to systematically bridging molecular interactions with macroscopic observables of electrochemical interfaces.

\section{Methods}

\subsection{Architecture of the electrostatic long-range MLIP}

For each atom $i$ in the system, the MLIP computes local invariant descriptors $B_i$ using the Cartesian atomic cluster expansion (CACE) formalism~\cite{cheng2024cartesian}, although we emphasize that any other short-ranged MLIP methods, based on either local atomic environment descriptors~\cite{drautz2019atomic,behler2007generalized,wang2018deepmd} or message-passing architectures~\cite{schutt2017schnet} can be used here.
These descriptors $B_i$ are mapped via a neural network to atomic energies $E_i$, which are then summed over all atoms in the system to obtain the total short-range energy
$E^{\mathrm{sr}} = \sum_{i=1}^N E_i$.

For the long-range electrostatic interactions, we use an extended Latent Ewald Summation (LES) method~\cite{Cheng2025Latent,kim2024learning,zhong2025machine,kim2025universal}:
the free charges on the atoms are explicitly considered, while the rapidly responding background electrons are treated as a dielectric medium with the high-frequency (electronic) relative permittivity $\varepsilon_\infty$. 
For each electrolyte atom $i$ (H, O, K, and F in this work), its latent charge $q^\mathrm{les}_{i \in \mathrm{electrolyte}}$ is predicted by its invariant $B_i$ features via a neural network. 
Such latent charges $q_i^\mathrm{les}$ are related to the free atomic charge $q_i$ by the constant scaling factor,
$q_i^\mathrm{les} = q_i / \sqrt{\varepsilon_\infty}$.
In this way, 
the electrostatic field produced by the free atomic charge $q_i$ of an atom $i$ is
\begin{equation} 
\boldsymbol{\mathcal{E}}_{i} (\mathbf{r}) 
= \dfrac{q_i}{4\pi \varepsilon_0 \varepsilon_\infty |\mathbf{r}-\mathbf{r}_i|^3} (\mathbf{r}-\mathbf{r}_i)
= \dfrac{q^\mathrm{les}_i}{4\pi \varepsilon_0 \sqrt{\varepsilon_\infty} |\mathbf{r}-\mathbf{r}_i|^3} (\mathbf{r}-\mathbf{r}_i), 
\label{eq:E}
\end{equation} 
with $\varepsilon_0$ being vacuum permittivity,
and the electric force (and energy) between a pair of atoms can be expressed without referring to $\varepsilon_\infty$, e.g.
\begin{equation}
    \mathcal{F}_{ij} = \dfrac{q_i^\mathrm{les} q_j^\mathrm{les}}{ 4\pi \varepsilon_0 r_{ij}^3} \mathbf{r}_{ij}.
    \label{eqn:Fij}
\end{equation}

The free charges on the Pt atoms in the metal electrode ($q_{i \in \mathrm{metal}}$) are then determined using the Siepmann-Sprik model~\cite{siepmann1995influence},
which, despite its simplicity and classic assumptions, has been shown to effectively describe polarization~\cite{grisafi2023predicting} and electron spillover at interfaces~\cite{Zhu2025Machine}.
In brief, the electrostatic term involving the electrode charges is written as
\begin{equation}
U_{\mathrm{elec}}(\mathbf{q}) = \frac{1}{2} \mathbf{q}^\mathrm{T} \mathbf{A} \mathbf{q} - \mathbf{B}^\mathrm{T} \mathbf{q} - \mathbf{v}^\mathrm{T} \mathbf{q},
\label{eq:Uelec}
\end{equation}
where $\mathbf{q} = (q_{1,2,\ldots \in \mathrm{metal}})^T$,
the vector $\mathbf{B}$ is the electrostatic potential due to the
electrolyte on each electrode atom (see Eqn.~\eqref{eq:E}),
and $\mathbf{v}$ is the applied external potential for each electrode atom. 
The symmetric matrix $\mathbf{A}$ depends on the positions
and the widths of the Gaussian charges of the electrode atoms,
and as the free charges are concentrated on the metal surface,
the entries are screened by a factor of $1/\varepsilon_\infty$ compared to bare charge interactions in vacuum.

$\mathbf{q}$ is solved by minimizing $U_{\mathrm{elec}}$ while maintaining the charge neutrality constraint for all the electrode atoms:
\begin{equation}
\mathbf{q} = \mathbf{S}^{-1} ( \mathbf{B} + \mathbf{v} ),
\label{eq:qstar}
\end{equation}
where 
\begin{equation}
\mathbf{S} \equiv \mathbf{A}^{-1} - \frac{\mathbf{A}^{-1} \mathbf{E} \mathbf{E}^\mathrm{T} \mathbf{A}^{-1}}
{\mathbf{E}^\mathrm{T} \mathbf{A}^{-1} \mathbf{E}},
\end{equation}
and $\mathbf{E}^\mathrm{T} = (1, \ldots, 1)$.
As fixed electrode atom positions are typically used in the Siepmann-Sprik model, the $\mathbf{S}^{-1}$ matrix can be pre-computed and stored, which substantially lowers the computational cost compared to the other charge equilibration schemes~\cite{rappe1991charge}.
The latent charges of the metal atoms are then determined by scaling the free charges, $q_i^\mathrm{les} = q_i / \sqrt{\varepsilon_\infty}$,
such that the expression for the electric force and energy (Eqn.~\eqref{eqn:Fij}) is maintained.

Indeed, in the absence of an external field ($\mathbf{v}=\mathbf{0}$), the constant factor $\varepsilon_\infty$ is inconsequential due to cancellations;
one can simply plug in $q^\mathrm{les}$ together with vacuum permittivity in solving Eqn.~\eqref{eq:Uelec}.
With an external field, the same procedure for solving $q^\mathrm{les}$ directly using Eqn.~\eqref{eq:Uelec} still applies, only that the magnitude of the $\mathbf{v}$ should be scaled by $\sqrt{\varepsilon_\infty}$. 
In this work, we take $\sqrt{\varepsilon_\infty}=1.33$ based on the experimental refractivity of water.

All these $q^\mathrm{les}$ charges are then used to compute
the long-range energy contribution $E^\mathrm{lr}$. 
For periodic systems, the Ewald summation is used:
\begin{equation}
E^\mathrm{lr} = \dfrac{1}{2\varepsilon_0V} \sum_{0<k<k_c} \dfrac{1}{k^2} e^{-\sigma^2 k^2/2} |S(\mathbf{k})|^2,
\label{eq:e_lr}
\end{equation}
with
\begin{equation}
S(\mathbf{k}) = \sum_{i=1}^N q_i^\mathrm{les} e^{i\mathbf{k}\cdot\mathbf{r}_i},
\label{eq:sfactor}
\end{equation}
where $\varepsilon_0$ is the vacuum permittivity, $\mathbf{k}$ is the reciprocal wave vector determined by the periodic cell of the system $L$, 
$V$ is the cell volume, and $\sigma$ is a smearing factor set to 1~\AA.

The total potential energy of the system is the sum of the short-range and the long-range contributions, $E=E^\mathrm{sr}+E^\mathrm{lr}$.
The forces are obtained through automatic differentiation of the total energy with respect to atomic positions.
The training of the MLIP is based on the conventional loss functions for energy and forces.

\subsection{Implementation and parameters of the MLIP}
The CACE code is implemented using PyTorch, and publicly available at
    \url{https://github.com/BingqingCheng/cace}.
We used the \texttt{Ewald} module for the LES part.
We also implemented a new \texttt{MetalWall} module for the Siepmann-Sprik method.
The addition of the external electric field to the system is also included in the \texttt{MetalWall} module.

The short-ranged CACE model uses a radial cutoff of $5.5$~\AA, 6 Bessel radial functions, $c = 12$, $\ell_\mathrm{max} = 3$, $\nu_\mathrm{max} = 3$, $N_\mathrm{embedding} = 5$, no message passing.
The long-range part based on LES uses a one-dimensional hidden variable, $\sigma = 1$~\AA{}, and $k_c = \pi$ ($\mathrm{dl}=2$~\AA). 
The Siepmann-Sprik model for metal uses a reciprocal width of electrode charge smearing of $1/1.805132~\mathrm{\AA}$.

\subsection{Finite field simulations}

The MD simulations using MLIP are performed in ASE using the CACE calculator.
The simulation cell is fully periodic in all three directions, 
with the two electrodes on either side of the box along the $z$ axis (without vacuum between them).
A potential difference $\Delta \Phi$ between the two electrodes is created by an electric field of
magnitude $\mathcal{E}=\Delta \Phi/l_z$ along the $z$ direction, where $l_z$ is the $z$-dimension of the periodic simulation cell.
This means, in determining the charges on the electrode atoms, $\mathbf{v}=\mathcal{E} \mathbf{z}$ in Eqn.~\eqref{eq:Uelec} and Eqn.~\eqref{eq:qstar}, where $\mathbf{z}$ contains the $z$-positions of each
electrode atom.
Meanwhile, the electric field acts on all electrolyte atoms via adding the term 
$E^\mathrm{ff} = \sum_{i\in \mathrm{electrolyte} }\mathcal{E}q_i$ to the total energy.

\subsection{DFT training set}

The DFT setup is the same as Ref.~\cite{Zhu2025Machine}:
The Perdew-Burke-Ernzerhof (PBE)~\cite{Perdew1996} functional with the Grimme D3 dispersion correction~\cite{grimme2010consistent}. 
All calculations were performed using the CP2K package~\cite{hutter2014cp2k}, under 3D periodic boundary conditions and with zero total charges.
The Gaussian basis sets were double zeta with one set of
polarization functions. 
The plane wave energy cutoff was set to 800~Ry.
All the DFT data in the training set are
generated without an explicit external bias, 
although in many configurations the electrolyte is highly polarized, which induces surface charge and therefore potential difference between the two electrode surfaces.

We fitted the first generation of the MLIP using 3687 training configurations (90\% train / 10\% valid split) from Ref.~\cite{Zhu2025Machine}, which contains configurations of the Pt(111)/aqueous KF electrolyte interfaces with various surface charge densities, but also some configurations of
the bulk water, bulk KF electrolyte, and the electrolyte/vacuum interface.
Another 1000 configurations from Ref.~\cite{Zhu2025Machine} are used as a hold-out set for evaluating the training errors.

We then iteratively expand the training set and improve the MLIP for a total of five more generations: in each round, we first collect configurations from simulations of the Pt(111)/aqueous KF electrolyte interfaces at different external fields and with different electrolyte concentrations, and bulk KF electrolyte at high concentrations.
We collected an additional 3580 structures, making the final training set contain 7267 configurations (95\% train / 5\% valid split).

\begin{table}[h]
\centering
\begin{tabular}{lcccc}
\hline\hline
  & E (train) & F (train) & E (test) & F (test)\\
\hline
DPLR &  1.305  & 75.00 & 0.9638 & 57.19 \\
Gen 1 & 0.29 & 33.9 & 0.47 & 34.6 \\
Gen 6 & 0.26 & 30.2 & 0.21 & 30.0 \\
\hline\hline
\end{tabular}
\caption{Root mean square errors (RMSEs) for the energies ($E$) in meV/atom and the atomic forces ($F$) in meV/\AA{} by comparing the results from the MLIP and DFT calculation, in either the training set or the testing set. The DPLR results are from Ref.~\cite{Zhu2025Machine}.}
\label{tab:rmse}
\end{table}

For testing the MLIPs, we collected configurations from the MLIP MD simulations of Pt(111)/water and Pt(111)/KF 2~M at 370~K, 
under different bias potentials ranging between 0 and 2.5~V. 
We recomputed the DFT energies and forces for these configurations and compared them with the MLIP predictions. 

In Table~\ref{tab:rmse} we compare the energy and force RMSEs for the training set (using the 1000 hold-out configurations) and the test set for the initial generation (Gen 1), and the last generation (Gen 6) MLIPs.
For comparison, we also included the errors from the DPLR model in Ref.~\cite{Zhu2025Machine}.
Note that the test set of the DPLR model has a different composition, although it was also collected  using a similar strategy from MLIP MD simulations at 330~K and at different surface charges.
Our Gen 6 model achieves improved accuracy compared to DPLR as well as the Gen 1 MLIP.
The production MD runs were all performed using the Gen 6 model.

\subsection{Benchmark of the MLIP on BEC}

\begin{figure}
\centering
\includegraphics[width=0.7\linewidth]{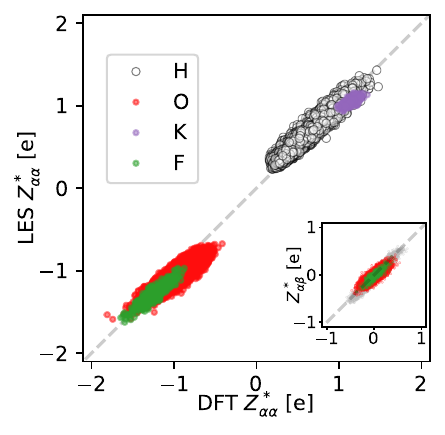}
     \caption{
     The comparison of the Born effective charge tensors ($Z^*$) computed from DFT and predicted using the LES.
     The main panels compare the diagonal elements of BEC ($Z^*_{\alpha\alpha}$), and the insets show the off-diagonal elements ($Z^*_{\alpha\beta}$ with $\alpha \ne \beta$).
    }
    \label{fig:bec}
\end{figure}

To validate whether our MLIPs can capture the correct electrostatics, we benchmark the predicted 
Born effective
charge tensors ($Z^*_{\alpha\beta}$), where $\alpha$ and $\beta$ indicate Cartesian directions, against density-functional perturbation theory (DFPT) calculations~\cite{gonze1997dynamical} in VASP~\cite{Kresse1993a,Kresse1994} using the PBE functional.
The BECs are not only sensitive indicators of the correct electrostatics, but are also crucial in atomistic simulations for electrical response properties.
In LES, the BEC for atoms in a homogeneous bulk system under PBCs is 
\begin{equation}
Z^*_{i \alpha\beta} 
= \Re\left[\exp(-ik r_{i\alpha}) \dfrac{\partial P_{\alpha} (k)}
{\partial r_{i\beta}}\right],
\label{eq:z-pbc}
\end{equation}
with $P_\alpha(k) 
    = \sum_{i=1}^{N} \sqrt{\varepsilon_\infty}\dfrac{q_i^\mathrm{les}}{ik} \exp(i k r_{i\alpha}) $.

We used a total of 50 configurations for the BEC comparison, each containing 64 water molecules and between 1-5 KF ion pairs (corresponding to molar concentrations between 1-4~M).
The high-frequency (electronic) relative permittivity $\varepsilon_\infty$ computed for these configurations ranges from 1.91 (1 KF pair in water) to 1.97 (5 KF pairs in water), while the $\varepsilon_\infty$ for water was computed to be 1.86. 
These values are quite consistent with $1.33^2=1.77$ from the experimental refractivity of water (1.33).

In Fig.~\ref{fig:bec} we compare the reference DFT BEC with the prediction using the MLIP, for both the diagonal and off-diagonal components. For all atomic elements, good agreement can be seen, suggesting that the MLIP can correctly capture the electrostatics of the electrolyte.

\subsection{Comparison of AIMD and MLIP MD}

In the \silabel, we show that the MLIP shows good agreement with previous AIMD simulations for the properties of bulk water and electrolyte solutions.
Moreover, MLIP MD at 370~K can well reproduce experimental water radial distribution functions, diffusion coefficients, and hydration shell properties for F$^-$ and K$^+$ ions.

\begin{figure}
\centering
\includegraphics[width=0.6\linewidth]{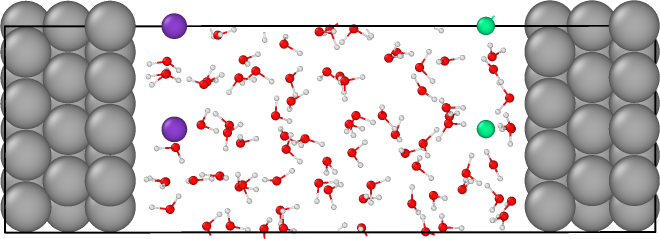}
\includegraphics[width=0.95\linewidth]{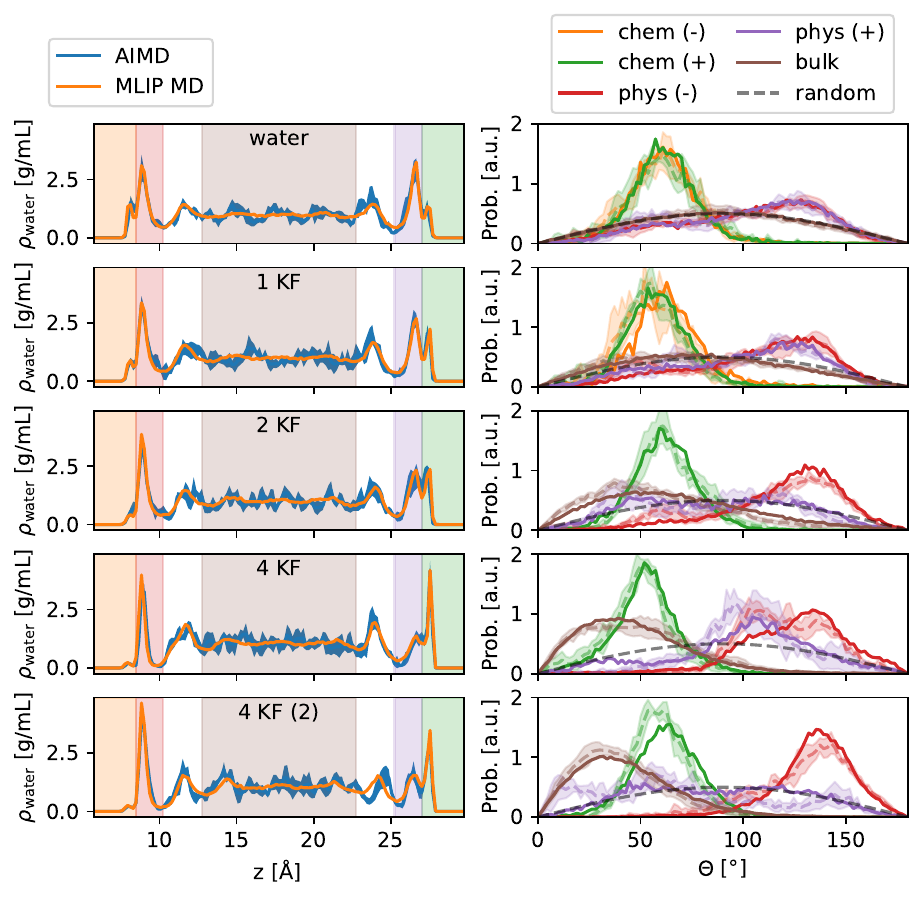}
     \caption{
     The comparison of water structure from AIMD and MLIP MD simulations for systems with different amounts of surface ions.
    \textbf{a}: A snapshot of an atomic configuration used in the AIMD and MLIP MD benchmark. 4 K ions and 4 F ions are fixed near the left and the right electrode, respectively.
    \textbf{b}: The water density profiles $\rho_\mathrm{water}$ as a function of the coordinate along the $z$-axis, perpendicular to the surface. 
    The numbers of ion pairs fixed near the electrodes are indicated.
    The orange, red, light brown, purple, and green shaded areas indicate the chemisorbed layer on cathode (chem ($-$)), physisorbed layer on cathode (phys ($-$)),
    chemisorbed layer on anode (chem ($+$)), physisorbed layer on anode (phys ($+$)), and bulk water, respectively.
    The 4~KF system has all four K ions and all four F ions in-plane, as illustrated in {a}. The 4~KF (2) set stacks two and two K/F ions on two $z$ planes.
    \textbf{c}: The orientation distributions of chemisorbed and physisorbed water molecules.
    The AIMD results are shown as dashed curves with shades, and the MLIP results are shown as solid curves.
    In \textbf{b} and \textbf{c}, the statistical uncertainties from the AIMD simulations are shown using the shaded regions.
    }
    \label{fig:aimd-mlpmd}
\end{figure}

To benchmark for the interfacial systems, we ran AIMD simulations using the same PBE-D3 DFT functional using CP2K, with the plane wave cutoff reduced to 400~Ry.
The Pt(111)/electrolyte interface system has 77 water molecules and 96 Pt atoms in a ($4\times4$) 6-layer slab, in a periodic simulation cell with dimension 
$(l_z =11.257~\mathrm{\AA},~l_y=9.749~\mathrm{\AA},~l_z=35.489~\mathrm{\AA})$.
Between 0-4 KF pairs are inserted close to the electrode surface on either side: K on the left and F on the right, as illustrated in Fig.~\ref{fig:aimd-mlpmd}a.
The simulation temperature was set at 420~K using the Nos\'e-Hoover thermostat, for fast equilibration.
All Pt atoms in the electrode and all K and F ions were frozen in their positions during the simulations.
No external field was applied; different
surface charge
states of the electrode were induced by the varying amount of the charged ions near the surface. 
The timestep was 1 fs. For each system, the production run was for 5~ps, after 2~ps of equilibration.

For the MLIP MD simulations on the same systems, a similar MD setup was used, except that the total simulation length was 1~ns including 200~ps of equilibration.

Fig.~\ref{fig:aimd-mlpmd} compares the water structure obtained from AIMD and MLIP MD simulations under different surface ion densities. 
The water density profiles are shown in Fig.~\ref{fig:aimd-mlpmd}b, with  chemisorbed, physisorbed, and bulk-like layers indicated. Fig.~\ref{fig:aimd-mlpmd}c shows the orientational distributions of water molecules in these layers.
Overall, the comparison demonstrates that the MLIP captures both the density and orientational ordering of interfacial water under varying ion conditions, consistent with AIMD benchmarks.

\subsection{MLIP MD simulation details}

The Pt(111)/water interface system has 233 water molecules and 216 Pt atoms in ($6\times6$) 6-layer slab, in a periodic simulation cell with dimension $l_x =16.869~\mathrm{\AA},~l_y=14.609~\mathrm{\AA},~l_z=41.478~\mathrm{\AA}$, without vacuum layers.
The Pt(111)/KF 2~M system has the same size and the Pt slab, but has 225 water molecules and 10 KF ion pairs.
The Pt(111)/KF 0.4~M system has 233 water molecules and 2 KF pairs.
The lattice constant of the Pt is 3.98~\AA. The electrolyte is filled between the space of 30~\AA{} between two Pt surfaces. 
The bulk water in the center of the simulation box has a density of about 1~g/mL.
We used the 1~fs timestep throughout. 
The simulation temperature was elevated to 370~K enforced using the Nos\'e-Hoover thermostat, as PBE water tends to be overstructured and has a high melting point of about 417~K~\cite{yoo2009phase}.
All Pt atoms in the electrode were frozen to their equilibrium lattice positions during the simulations.
In the finite field simulations (as the setup described above), a constant electric field of
magnitude $-\Delta \Phi/l_z$ along the $z$ direction was applied to all atoms in the cell, creating a potential drop of $\Delta \Phi$ between the two electrodes on either side of the simulation cell.
For each system, we ran independent simulations at a range of $\Delta \Phi$ values:
0.00,
0.08,
0.15,
0.23,
0.30,
0.38,
0.53,
0.75,
0.90,
1.13,
1.28,
1.50,
1.65,
1.88,
2.03,
2.26,
2.41,
and
2.63~V.
For Pt(111)/water,  each run lasted 1~ns, except for the higher voltage cases where water dissociations happened 3-5~ns was used. 
For the Pt(111)/KF 2~M and Pt(111)/KF 0.4~M systems, each run lasted 3-5~ns.
Snapshots per 1~ps from the MD trajectories were collected for post-processing.
We note that such simulation times are not sufficient to ergodically sample anodic water dissociation and the equilibrium concentration of dissociated species.
However, using enhanced sampling for reaction barriers in this case is not straightforward due to the multiple steps involved in the dissociation and the subsequent water diffusion.

\subsection{Details on data analysis}

For estimating $\avg{Q}$ efficiently utilizing the finite-field MD simulations at different values of $\Delta\Phi_j$, we used a reweighting scheme:
\begin{equation}
    \avg{Q}(\Delta \Phi) = 
    \sum_j w_j 
    \dfrac{\avg{Q e^{\beta Q(\Delta \Phi - \Delta \Phi_j)}}_{\Delta \Phi_j}}
    {\avg{e^{\beta Q(\Delta \Phi - \Delta \Phi_j)}}_{\Delta \Phi_j}}
    \label{eq:reweight}
\end{equation}
where $\avg{\ldots}_{\Delta \Phi_j}$ denotes the ensemble average sampled from MD simulations at the bias $\Delta\Phi_j$,
and $\sum_j w_j =1$ is the set of optimal weights that minimize the uncertainty of the final $\avg{Q}$ estimates~\cite{shirts2008statistically}.

For classifying the species (see the legends in Fig.~\ref{fig:reaction}a,b) involved in a snapshot from the MD trajectories,
we first built a neighborlist using the following cutoffs:
$r_\mathrm{cut}(\mathrm{O-H}) = r_\mathrm{cut}(\mathrm{H-H}) = 1.25$~\AA, $r_\mathrm{cut}(\mathrm{H-F}) = r_\mathrm{cut}(\mathrm{O-F}) = 1.15$~\AA.
If two atoms are within such cutoffs, they belong to the same species.

For computing the proton hopping statistics, we used three different algorithms for identifying individual hopping events:
The first is to label each H atom that has a change of its nearest heavy atom neighbor between the previous and the current snapshot.
The second is to find all the species involved using a network analysis based on the neighborlist, and then identify the reaction by expanding over all species in the previous and current snapshots that have changed atomic constituents. 
The third is to first tag positively charged or negatively charged species in each frame, and then find the reactions related to them between the previous and the current frame.
The code for the three algorithms is provided in the \silabel.
They differ mostly by the treatment of transient events, e.g.
a H$_3$O$^+$ forms a Zundel complex with a H$_2$O at one step and returns to its former state in the next.
In the analysis presented in Fig.~\ref{fig:reaction}, the second algorithm is used, and such transient events are not counted.
Nevertheless, all three algorithms yield less than 10\% difference in the number of hopping events identified.

The red and blue curves Fig.~\ref{fig:reaction}d show the Poisson model prediction for hopping statistics: 
For a time window $T=1$~ps, the probability of observing $k$ hops is:
\begin{equation}
    P[k] = \exp \left(-\frac{T}{\tau}\right)\frac{(T/\tau)^k}{k!},
\end{equation}
where $\tau$ is from the average residence time.

\subsection{Benchmark on classical simulations}
We benchmark the MLIP architecture based on a classical empirical force field with fixed charges.
The ground truth is an SPC/Fw water model with fixed charges interacting with Pt(111) electrode with flexible Gaussian charges determined by the Siepmann-Sprik model~\cite{siepmann1995influence},
as well as Lennard-Jones interactions between water and the metal.
This exercise serves three purposes: validate the implementation, demonstrate that the charges of both electrolyte and electrode can be inferred solely from energies and forces, and test the generalization of the model across temperatures and external potentials.

\begin{figure}
  \centering
    \subfigimg[width=0.5\linewidth]{\figLabel{a}}{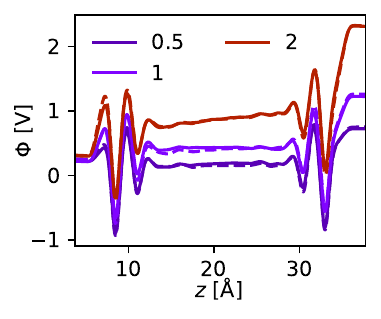} 
    \subfigimg[width=0.42\linewidth]{\figLabel{b}}{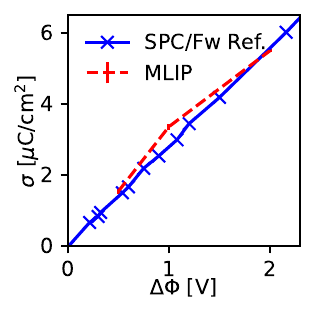} 
  \caption{
  The comparison between the classical empirical force field (SPC/Fw water and Siepmann-Sprik model for Pt) and the MLIP trained based on the classical reference, for the Pt(111)/water system at 420~K under a range of bias potential $\Delta\Phi$.  \\
  \textbf{a}: Electric potential $\Phi$ along the $z$-axis.
\textbf{b}: Area–specific charge $\sigma=\avg{Q}/S$.
The dashed lines are for the MLIP and the solid lines are for the classical force field.
  }
    \label{fig:class}
\end{figure}

The training set contains the energies and forces of 2814 configurations collected from equilibrium MD simulations under zero external field. 
The training errors are tiny: 0.14~meV/atom in energy RMSE and 8.8~meV/\AA{} in force RMSE.
We then compared the finite-field MD simulations of the Pt(111)/water interface using the classical force field and using the trained MLIP.
The MD temperature was set at 420~K, in order to test the generalization of the MLIP at an elevated temperature.
Fig.~\ref{fig:class} shows that the electric potential and surface charge density for the Pt(111)/water system described by the classical reference and the MLIP trained on it have excellent agreement.

\subsection{Reference classical simulations}
Using the same SPC/Fw water model and the Siepmann-Sprik metal electrode as described above,
we performed finite-field simulations  using the ELECTRODE package~\cite{ahrens2022electrode} in LAMMPS~\cite{plim95jcp}.
The same size and composition for the Pt(111)/water interface system as the MLIP MD simulation was used in the classical MD.
1~fs timestep was used and the total simulation time was 2~ns at each $\Delta \Phi$. 
We performed two sets of NVT simulations at different temperatures using the Nos\'e-Hoover thermostat:
the 370~K surface charge and capacitance results are reported in Fig.~\ref{fig:C}, and the 420~K results are in Fig.~\ref{fig:class}.

\textbf{Acknowledgements}
B.C. is deeply grateful to David Limmer and Shannon Boettcher for valuable discussions and feedback on the manuscript, and for teaching me something about electrochemistry.

\textbf{Author contributions}
X.W. benchmarked the MLIP on bulk systems, helped analyze reaction dynamics, and prepared the SI.
J.C. prepared the LAMMPS input files for molecular dynamics simulations using the classical forcefield.
Z.Z. tested the CP2K input files.
F.S. advised on the CP2K settings.
J.L. benchmarked the MLIP on liquid-vapor interfacial systems.
B.C. designed and performed the research, and wrote the paper.
All authors reviewed the paper.

\textbf{Code availability}~The CACE package is publicly available at \url{https://github.com/BingqingCheng/cace}.

\textbf{Data availability statement}
Training data, trained MLIP, and data analysis scripts
generated for the study are in the
SI repository \url{https://github.com/BingqingCheng/pt-electrolyte}.

\textbf{Competing Interests}~
B.C. has an equity stake in AIMATX Inc.
University of California, Berkeley has filed a provisional patent for the Latent Ewald Summation algorithm.

\end{document}